# Macroscopic Quantum Electrodynamics with Gain: Modified Fluctuations and Their Consequences

**Daigo Oue**[1,2,3]

[1] RIKEN Centre for Advanced Photonics, RIKEN, Saitama 351-0198, Japan
[2] Instituto Superior Técnico—University of Lisbon and Instituto de Telecomunicações, Lisbon 1049-001, Portugal
[3] The Blackett Laboratory, Imperial College London, London SW7 2AZ, United Kingdom

E-mail: daigo.oue@gmail.com

### Abstract

Macroscopic quantum electrodynamics (MQED) provides a unified framework for describing quantum electromagnetic fields in arbitrary macroscopic environments. Central to this theory is the field correlation, which governs both radiative (e.g., Lamb shifts and the Purcell effect) and mechanical phenomena, such as van der Waals and Casimir forces. In this tutorial, we provide an overview of MQED and its extension to active media, highlighting fluctuation-induced forces as manifestations of gain-modified field correlations.

### Glossary

*Macroscopic quantum electrodynamics; Lamb shift; Purcell effect; Casimir forces*

### Introduction

Electromagnetic fluctuations are an unavoidable feature of both quantum and thermal physics. Even in a vacuum, fluctuating electromagnetic (EM) fields give rise to observable effects such as spontaneous emission, the Lamb shift, and dispersion forces. These are traditionally studied in different subfields—quantum optics on one hand [1,2] and Casimir physics on the other [3–5]; however, they share a common origin: the correlations of field fluctuations.

Fluctuational electrodynamics, also known as macroscopic quantum electrodynamics (MQED), draws inspiration from electromagnetism of continuous media [5–8] and provides a unified theoretical framework for describing these effects in realistic material environments, which can be inhomogeneous, dispersive, and absorptive. The framework has been extended to include optical gain [9], enabling the study of EM fluctuations in active media, such as spontaneous emission and atom population dynamics [10–14] and Casimir-type effects [15–19].

Active environments are increasingly relevant in nanophotonics and metamaterials, where optical gain can compensate for losses. While most studies focus on radiative phenomena such as spontaneous emission [10–14], the mechanical consequences of modified field fluctuations remain less explored. In passive systems, field fluctuations lead to van der Waals and Casimir forces. When motion or a static electric bias is introduced, the system becomes effectively active and may experience longitudinal (frictional) [16] or transverse (Hall-like) forces [17] originating from asymmetries in the underlying field fluctuations.

The objective of this tutorial is to introduce the MQED framework and its extension to active systems. We aim to (i) explain how radiative and mechanical effects originate from the field correlations, (ii) clarify how optical gain can be incorporated consistently into MQED, and (iii) outline how this framework predicts unconventional, longitudinal (frictional) and transverse (Hall-like) forces.

**Phenomenological field quantisation in passive dielectrics**

We begin at the level of equations of motion. The macroscopic Maxwell equations are interpreted as emerging from underlying Heisenberg–Langevin equations [1,20,21]. Our goal is to construct field operators that are consistent with the macroscopic response of matter and the canonical commutation relations.

In the frequency domain, the macroscopic Maxwell equations can be reduced to

$$\nabla \times \nabla \times \boldsymbol{E}^{-}(\boldsymbol{r},\omega) - \frac{\omega^2}{c^2}\epsilon(\boldsymbol{r},\omega)\boldsymbol{E}^{-}(\boldsymbol{r},\omega) = 0, \tag{1}$$

where the superscript represents the signature in the harmonic time dependence [i.e., the positive-frequency field evolves as $\boldsymbol{E}^{-}(\boldsymbol{r},\omega,t) = \boldsymbol{E}^{-}(\boldsymbol{r},\omega)e^{-i\omega t}$]. Note that the magnetic response is neglected for simplicity ($\mu_{\mathrm{r}} = 1$). Here, we have the frequency-dependent response function $\epsilon(\boldsymbol{r},\omega) = \sum_{\ell=1}^{N}\epsilon_\ell(\boldsymbol{r},\omega)$, where $N$ denotes the number of physically distinct channels contributing to the dielectric response [$\epsilon_\ell(\boldsymbol{r},\omega)$ describes the response in the $\ell$th channel]. Its real $\epsilon'_\ell$ and imaginary parts $\epsilon''_\ell$ correspond to dispersion (retardation) and dissipation (energy exchange), implying that the macroscopic response arises from underlying dynamical degrees of freedom [22–24].

The existence of underlying variables suggests that the material must be described by equations of motion. Let us restrict ourselves to the linear-response regime, where deviations in the material subsystem around an equilibrium are well described by harmonic oscillators. The harmonic oscillator is the simplest linearly responding system. Indeed, many elementary excitations admit such a description (e.g., plasmons, excitons, and magnons).

It is thus natural to model the material as a collection of harmonic oscillators labelled by position and frequency in each channel: We consider a vector-valued, bosonic annihilation operator $\hat{\boldsymbol{a}}_\ell(\boldsymbol{r},\omega)$, evolving as $\hat{\boldsymbol{a}}_\ell(\boldsymbol{r},\omega,t) = \hat{\boldsymbol{a}}_\ell(\boldsymbol{r},\omega)e^{-i\omega t}$ and satisfying $\left[\hat{a}_{\ell_1,i_1}(\boldsymbol{r},\omega), \hat{a}^{\dagger}_{\ell_2,i_1}(\boldsymbol{r}',\omega')\right] = \delta_{\ell_1\ell_2}\delta_{i_1i_2}\delta_{\boldsymbol{r},\boldsymbol{r}'}\delta_{\omega,\omega'}$. Note that a shorthand notation for the Dirac delta function is applied [e.g., $\delta_{\boldsymbol{r},\boldsymbol{r}'} \equiv \delta(\boldsymbol{r}-\boldsymbol{r}')$]. The field-oscillator coupling strength is not yet specified: The response function determines the macroscopic field behaviour, but it does not uniquely fix how the underlying oscillators contribute to the current density. This coupling is determined by imposing additional consistency on the quantised theory.

We now promote the classical description to an operator framework. The macroscopic Maxwell equation (1) describes expectation values of the field. At the operator level, the dynamics must be governed by the Heisenberg equations. In the presence of dissipation, they necessarily include Langevin noise terms. The quantum macroscopic Maxwell equation in the frequency domain may therefore be written as

$$\nabla \times \nabla \times \hat{\boldsymbol{E}}^{-}(\boldsymbol{r},\omega) - \frac{\omega^2}{c^2}\epsilon(\boldsymbol{r},\omega)\hat{\boldsymbol{E}}^{-}(\boldsymbol{r},\omega) = i\omega\mu_0\hat{\boldsymbol{j}}^{-}_{\mathrm{N}}(\boldsymbol{r},\omega), \tag{2}$$

where $\hat{\boldsymbol{j}}^{-}_{\mathrm{N}}(r,\omega)$ is understood as an operator-valued current density associated with fluctuations in the material. The field operator evolves as $\hat{\boldsymbol{E}}^{-}(\boldsymbol{r},\omega,t) = \hat{\boldsymbol{E}}^{-}(\boldsymbol{r},\omega)e^{-i\omega t}$, and the actual field should be $\hat{\boldsymbol{E}}(\boldsymbol{r},\omega) = \hat{\boldsymbol{E}}^{-}(\boldsymbol{r},\omega) + \hat{\boldsymbol{E}}^{+}(\boldsymbol{r},\omega)$, where the negative-frequency part is given by $\hat{\boldsymbol{E}}^{+}(\boldsymbol{r},\omega) = [\hat{\boldsymbol{E}}^{-}(\boldsymbol{r},\omega)]^{\dagger}$. The presence of the Langevin noise current $\hat{\boldsymbol{j}}^{-}_{\mathrm{N}}$ dissipates the field, implying that its action is annihilation-like. More precisely, the positive-frequency component of the noise current operator must remove electromagnetic excitations. This observation motivates us to write the noise current $\hat{\boldsymbol{j}}^{-}_{\mathrm{N}}$ in terms of the annihilation operators $\hat{\boldsymbol{a}}_\ell(\boldsymbol{r},\omega)$ associated with the underlying oscillators. We therefore assume $\hat{\boldsymbol{j}}^{-}_{\mathrm{N}}(\boldsymbol{r},\omega) = \sum_{\ell=1}^{N}C_\ell(\boldsymbol{r},\omega)\hat{\boldsymbol{a}}_\ell(\boldsymbol{r},\omega)$ with an unknown coefficient $C_\ell(\boldsymbol{r},\omega)$. Its explicit form cannot be determined from classical macroscopic electrodynamics alone and must instead be fixed by imposing consistency conditions on the field operators.

The operator equation (2) can be solved using the Green tensor $\boldsymbol{G}(\boldsymbol{r},\boldsymbol{r}',\boldsymbol{\omega})$ [25], defined as the solution of $[\boldsymbol{\nabla}\times\boldsymbol{\nabla}\times -(\omega^2/c^2)\epsilon]G = I\delta_{\boldsymbol{r},\boldsymbol{r}'}$, where $I$ is the unit tensor, and we used the shorthand notation $\delta_{\boldsymbol{r},\boldsymbol{r}'} \equiv \delta(\boldsymbol{r} -$ $\boldsymbol{r}'$). With this Green tensor, the electric-field and vector-potential operators in the frequency domain can be written as

$$\widehat{\boldsymbol{E}}^{-}(\boldsymbol{r},\omega) = i\omega\mu_0 \int G(\boldsymbol{r},\boldsymbol{r}',\omega)\cdot \hat{\boldsymbol{j}}_{\mathrm{N}}^{-}(\boldsymbol{r}',\omega)\,\mathrm{d}\boldsymbol{r}', \qquad \widehat{\boldsymbol{A}}^{-}(\boldsymbol{r},\omega) = \mu_0 \int G(\boldsymbol{r},\boldsymbol{r}',\omega)\cdot \hat{\boldsymbol{j}}_{\mathrm{N}}^{-}(\boldsymbol{r}',\omega)\,\mathrm{d}\boldsymbol{r}', \tag{3}$$

where $\boldsymbol{\mu_0}$ is the vacuum permeability. Remember that the actual vector potential is $\widehat{\boldsymbol{A}}(\boldsymbol{r},\omega) = \widehat{\boldsymbol{A}}^{-}(\boldsymbol{r},\omega) + \widehat{\boldsymbol{A}}^{+}(\boldsymbol{r},\omega)$ with $\widehat{\boldsymbol{A}}^{+}(\boldsymbol{r},\omega) = [\widehat{\boldsymbol{A}}^{-}(\boldsymbol{r},\omega)]^{\dagger}$. These expressions highlight a central feature of the MQED framework: all geometric and material information is contained in the Green tensor, while the field's quantum properties enter through the operator structure of the noise current. The problem of field quantisation in macroscopic media is therefore reduced to determining the correct operator structure of $\hat{\boldsymbol{j}}_{\mathrm{N}}^{-}$.

To fix it, we impose the canonical commutation relation, $\left[\widehat{\boldsymbol{E}}(\boldsymbol{r},\omega),\widehat{\boldsymbol{A}}(\boldsymbol{r}',\omega)\right] = i\hbar I\delta_{\boldsymbol{r},\boldsymbol{r}'}$. Substituting the Green-function representation (3) into this commutation relation yields a constraint on the coupling kernel $C_\ell(\boldsymbol{r},\omega)$. One finds that consistency with the canonical commutation relations requires the noise current operator to be proportional to the square root of the imaginary part of the permittivity [26],

$$\hat{\boldsymbol{j}}_{\mathrm{N}}^{-}(\boldsymbol{r},\omega) = \frac{\omega}{c}\sqrt{\frac{\hbar}{\pi\mu_0}}\sum_{\ell=1}^{N}\sqrt{\epsilon_\ell''(\boldsymbol{r},\omega)}\;\hat{\boldsymbol{a}}_\ell(\boldsymbol{r},\omega). \tag{4}$$

Once the structure of the noise current operator is fixed, its statistical properties follow directly:

$$\langle \hat{\boldsymbol{j}}_{\mathrm{N}}^{-}(\boldsymbol{r},\omega)\hat{\boldsymbol{j}}_{\mathrm{N}}^{+}(\boldsymbol{r}',\omega)\rangle = \frac{\hbar}{\pi\mu_0}\frac{\omega^2}{c^2}\epsilon''(\boldsymbol{r},\omega)I\delta_{\boldsymbol{r},\boldsymbol{r}'}, \qquad \langle \hat{\boldsymbol{j}}_{\mathrm{N}}^{+}(\boldsymbol{r},\omega)\hat{\boldsymbol{j}}_{\mathrm{N}}^{-}(\boldsymbol{r}',\omega)\rangle = 0, \tag{5}$$

where $\hat{\boldsymbol{j}}_{\mathrm{N}}^{+}(\boldsymbol{r}',\omega') = [\hat{\boldsymbol{j}}_{\mathrm{N}}^{-}(\boldsymbol{r},\omega)]^{\dagger}$, and the expectation value is taken with respect to the vacuum state (i.e., $\langle \dots \rangle \equiv \langle 0| \dots |0\rangle$). The relation (5) can be regarded as a fluctuation-dissipation relation, since the correlation function is written in terms of the imaginary part of the response function. The choice of the current structure (4) that guarantees the canonical commutation relations also leads to the current-current correlations (5) consistent with the fluctuation-dissipation theorem. Moreover, the MQED approach developed here is fully consistent with microscopic quantisation schemes, such as those in Ref.[22]. The consistency has been carefully studied [23,27,28]. It has also been verified that the MQED formalism in the lossless limit reduces to the conventional mode-expansion method [29]. Furthermore, the framework can be extended to rather general passive media, including materials with dielectric and magnetic responses, magnetoelectric coupling, and nonlocal or nonreciprocal behaviour (see, e.g., Refs. [24,26,33]).

**Field Correlation Functions**

The field operator in MQED is fully determined by the classical Green's tensor and the noise current. The field-field correlation, playing a central role in the quantum theory, can be written as

$$\langle \widehat{\boldsymbol{E}}(\boldsymbol{r},\omega)\widehat{\boldsymbol{E}}(\boldsymbol{r}',\omega)\rangle = \frac{\hbar}{\pi\epsilon_0}\int \bar{G}(\boldsymbol{r},\boldsymbol{s},\omega)\cdot\epsilon''(\boldsymbol{s},\omega)\cdot\bar{G}^{\dagger}(\boldsymbol{r}',\boldsymbol{s},\omega)\,\mathrm{d}\boldsymbol{s} = \frac{\hbar}{\pi\epsilon_0}\Im\{\bar{G}(\boldsymbol{r},\boldsymbol{r}')\}, \tag{6}$$

where we defined $\bar{G}(\boldsymbol{r},\boldsymbol{r}') = (\omega^2/c^2)G(\boldsymbol{r},\boldsymbol{r}')$. Note that the operators appearing in Eq. (6), $\widehat{\boldsymbol{E}}(\boldsymbol{r},\omega)$ and $\widehat{\boldsymbol{E}}(\boldsymbol{r}',\omega)$, correspond to the electric-field amplitudes evaluated at $\boldsymbol{r}$ and $\boldsymbol{r}'$, respectively. In Eq. (6), we used a well-known identity $\Im\{\bar{G}(\boldsymbol{r},\boldsymbol{r}')\} = \int \bar{G}(\boldsymbol{r},\boldsymbol{s})\epsilon(\boldsymbol{s})\bar{G}^{\dagger}(\boldsymbol{r}',\boldsymbol{s})\mathrm{d}\boldsymbol{s}$ (see, e.g., Ref. [26] for the proof).

Radiative observables provide familiar examples: The spontaneous decay rate $\Gamma$ of an emitter located at $\boldsymbol{r}_0$ with a transition frequency $\omega_0$ and a dipole moment $\boldsymbol{d}_{\mathrm{e}}$ can be written in terms of the field correlator [27]:

$$\Gamma = \frac{\boldsymbol{d}_{\mathbf{e}}}{\hbar}\cdot 2\pi\langle \widehat{\boldsymbol{E}}(\boldsymbol{r}_0,\omega_0)\widehat{\boldsymbol{E}}(\boldsymbol{r}_0,\omega_0)\rangle\cdot\frac{\boldsymbol{d}_{\mathbf{e}}}{\hbar}. \tag{7}$$

The Purcell factor, describing the enhancement of the spontaneous emission rate, can be obtained by comparing $\Gamma$ with the free-space counterpart [13]. Likewise, the emitter's energy-level shifts (Lamb shifts) due to the interaction with the vacuum depend on the field correlation,

$$\delta\omega = \frac{\boldsymbol{d}_{\mathbf{e}}}{\hbar}\cdot\left(\int\frac{\langle\widehat{\boldsymbol{E}}(\boldsymbol{r}_0,\omega)\widehat{\boldsymbol{E}}(\boldsymbol{r}_0,\omega)\rangle}{\omega_0-\omega}\mathrm{d}\boldsymbol{\omega}\right)\cdot\frac{\boldsymbol{d}_{\mathbf{e}}}{\hbar}, \tag{8}$$

where the principal value should be taken. Thus, radiative properties are governed by the field correlator.

When the macroscopic environment is inhomogeneous, the level shift becomes position-dependent. The derivative is finite, producing a (typically attractive) force on the emitter. This is the origin of the van der Waals and Casimir–Polder forces (see, e.g., Refs. [30–32]). Moreover, the expectation value of the Maxwell stress tensor, $T_{ij} = T_{ij}^E + T_{ij}^B$, is fully determined by the field correlation function:

$$T_{ij}^O \equiv \epsilon_0\left(\langle\{\hat{O}_i(\boldsymbol{r})\hat{O}_j(\boldsymbol{r}')\}\rangle - \frac{1}{2}\delta_{ij}\mathrm{Tr}\langle\{\hat{O}_i(\boldsymbol{r})\hat{O}_j(\boldsymbol{r}')\}\rangle\right)_{\boldsymbol{r}'=\boldsymbol{r}} \quad (O = E, B), \tag{9}$$

where we should symmetrise the product of operators to respect the noncommutativity (e.g., $\langle\{\hat{E}_i(\boldsymbol{r},\omega)\hat{E}_j(\boldsymbol{r}',\omega)\}\rangle \equiv \langle\hat{E}_i(\boldsymbol{r},\omega)\hat{E}_j(\boldsymbol{r}',\omega)\rangle/2 + \langle\hat{E}_j(\boldsymbol{r}',\omega)\hat{E}_i(\boldsymbol{r},\omega)\rangle/2$). Note that the magnetic-field operator is defined as $\hat{B}_i(\boldsymbol{r}) = \sum_{jk}\varepsilon_{ijk}\partial_{r_j}\hat{A}_k(\boldsymbol{r})$, where $\varepsilon_{ijk}$ is the Levi–Civita symbol, and we wrote $\partial_{r_j} \equiv \partial/\partial r_j$. After some algebra, we can write the magnetic field correlation function in terms of the electric field correlation,

$$\langle\{\hat{B}_i(\boldsymbol{r})\hat{B}_j(\boldsymbol{r}')\}\rangle = \sum_{k,\ell,m,n}\boldsymbol{\varepsilon_{ik\ell}}\frac{\partial_{r_k}}{\omega}\varepsilon_{jmn}\frac{\partial_{r'_m}}{\omega}\langle\{\hat{E}_\ell(\boldsymbol{r})\hat{E}_n(\boldsymbol{r}')\}\rangle\,. \tag{10}$$

Therefore, the correlator (6) admits a direct mechanical interpretation. In particular, the shear component ($i \neq j$) is directly linked to the field correlation function.

Radiative and mechanical phenomena, therefore, represent different observables derived from the same fluctuations. The distinction lies not in their origin, but in how the correlation function is contracted with dipole moments or incorporated into the stress tensor. This unified viewpoint (see Figure 1) still holds even when extended to active systems. Once the correlation function is determined, both radiative and mechanical effects follow systematically.

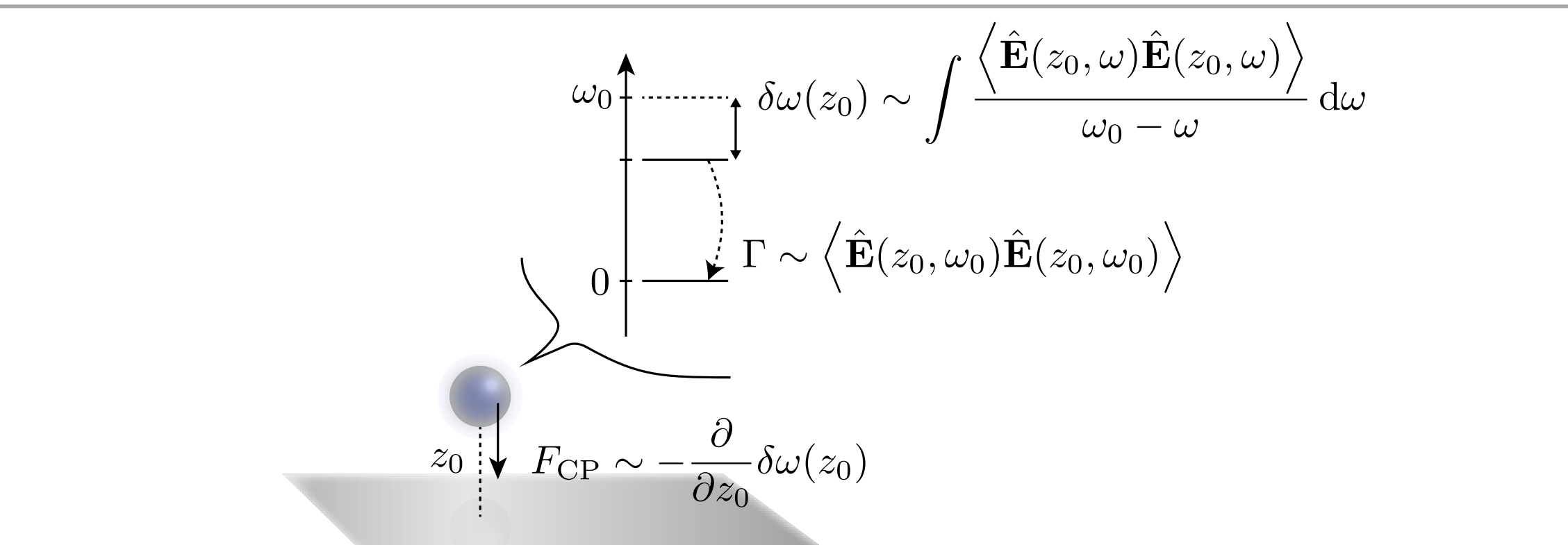

**Figure 1.** The spontaneous emission rate $\Gamma$ and energy-level shift $\delta\omega(z_0)$ of an emitter are affected by the inhomogeneous environment (single surface in this example; the surface-emitter distance: $z_0$). The position-dependent level shift may lead to the fluctuation-induced (Casimir-Polder) force, $F_{\mathrm{CP}} \equiv -\partial(\delta\omega)/\partial z_0$. All these quantities are described in a unified manner using the field correlation.

### Optical Gain and the Modified Fluctuation–Dissipation Relation

In Section 2, we considered passive media, where the imaginary parts of response functions are non-negative. In active media, however, the situation changes qualitatively. In some channels, the imaginary parts of their response functions become negative, indicating optical gain; energy is transferred from the material to electromagnetic subsystems. Let us split the loss and gain contributions,

$$\epsilon''(\boldsymbol{r},\omega) = \epsilon''_{>}(\boldsymbol{r},\omega) + \epsilon''_{<}(\boldsymbol{r},\omega), \qquad \epsilon''_{>}(\boldsymbol{r},\omega) = \sum_{\epsilon''_\ell>0} \epsilon''_\ell(\boldsymbol{r},\omega)\,, \qquad \epsilon''_{<}(\boldsymbol{r},\omega) = \sum_{\epsilon''_\ell<0} \epsilon''_\ell(\boldsymbol{r},\omega)\,. \tag{11}$$

At first sight, $\epsilon''_\ell < 0$ seems problematic and challenges the fluctuation–dissipation relation, since the $\ell$th contribution to the noise current was found to be proportional to $\sqrt{\epsilon''_\ell}$. Early discussions raised concerns about whether the standard macroscopic quantisation procedure could be consistently extended to active media [9,15,35–39]. The resolution lies in recognising that the fluctuation–dissipation relation is not an independent postulate, but a consequence of the underlying oscillator model and its quantum statistics. It can be shown [6][9] that an appropriate combination of operators,

$$\hat{\boldsymbol{j}}_{\mathrm{N}}^{-}(\boldsymbol{r},\omega) = \frac{\omega}{c}\sqrt{\frac{\hbar}{\pi\mu_0}}\sum_{\epsilon''_\ell>0}\sqrt{\epsilon''_\ell(\boldsymbol{r},\omega)}\,\hat{\boldsymbol{a}}_\ell(\boldsymbol{r},\omega) + \frac{\omega}{c}\sqrt{\frac{\hbar}{\pi\mu_0}}\sum_{\epsilon''_\ell<0}\sqrt{-\epsilon''_\ell(\boldsymbol{r},\omega)}\,\hat{\boldsymbol{b}}^{\dagger}_\ell(\boldsymbol{r},\omega)\,, \tag{12}$$

recovers the canonical commutation relation. It is important to note that the correct commutation relation can be obtained only if all poles of the Green's function are in the lower half of the complex frequency plane [1]. In other words, the system should be stable to use the MQED framework. The assignment (particularly, the second term) in Eq. $(12)$ can be interpreted as follows: In a channel providing gain, the presence of current may amplify (rather than dissipate) the field, implying that its action is creation-like (not annihilation-like). It is worth noting that the structure of the field operator, $\hat{\boldsymbol{E}}^{-}(\boldsymbol{r}) = i\omega\mu_0\int G(\boldsymbol{r},\boldsymbol{r}')\cdot\hat{\boldsymbol{j}}_{\mathrm{N}}^{-}(\boldsymbol{r}')\mathrm{d}\boldsymbol{r}'$, remains unchanged; the noise current changes. Thus, even in the presence of optical gain, expressions developed in the MQED framework [e.g., Eqs. $(7)$ and $(8)$] are still valid. We should note that the current correlation functions are now modified,

$$\langle\hat{\boldsymbol{j}}_{\mathrm{N}}^{-}(\boldsymbol{r},\omega)\hat{\boldsymbol{j}}_{\mathrm{N}}^{+}(\boldsymbol{r}',\omega)\rangle = \frac{\hbar}{\pi\mu_0}\frac{\omega^2}{c^2}\epsilon''_{>}(\boldsymbol{r},\omega)I\delta_{\boldsymbol{r},\boldsymbol{r}'}, \qquad \langle\hat{\boldsymbol{j}}_{\mathrm{N}}^{+}(\boldsymbol{r}',\omega)\hat{\boldsymbol{j}}_{\mathrm{N}}^{-}(\boldsymbol{r},\omega)\rangle = \frac{\hbar}{\pi\mu_0}\frac{\omega^2}{c^2}|\epsilon''_{<}(\boldsymbol{r},\omega)|I\delta_{\boldsymbol{r},\boldsymbol{r}'}, \tag{13}$$

The first (second) correlator represents the fluctuations in the loss (gain) channel. The second contribution concerns the origin of unconventional fluctuation-induced forces.

**Unconventional Fluctuation-Induced Forces with Optical Gain**

In passive systems, dispersion forces are typically attractive and determined by equilibrium fluctuations. In contrast, active media can alter the direction of the interaction. For instance, optical gain may lead to repulsive Casimir forces [7,10,11]. Moreover, optical gain may generate qualitatively new fluctuation-induced forces:

(i) *Drag (longitudinal) Forces*. —When a macroscopic body moves relative to another with a constant velocity $\boldsymbol{v}$, the body's response is modified. Even if the body is passive at rest [i.e., the response satisfies $\mathrm{sgn}\{\epsilon''(\omega)\} = \mathrm{sgn}(\omega) > 0\ (\omega > 0)$], it provides direction-dependent optical gain when moving: the response is Doppler-shifted, $\omega \to \omega - \boldsymbol{k}\cdot\boldsymbol{v}$, and we have the negative imaginary part in the response function, $\mathrm{sgn}\{\epsilon''_{\mathrm{mov}}(\omega)\} = \mathrm{sgn}\{\epsilon''(\omega - \boldsymbol{k}\cdot\boldsymbol{v})\} < 0\ (0 < \omega < \boldsymbol{k}\cdot\boldsymbol{v})$.

The gain-induced current fluctuation $\langle\hat{\boldsymbol{j}}_{\mathrm{N}}^{+}(\boldsymbol{r}',\omega)\hat{\boldsymbol{j}}_{\mathrm{N}}^{-}(\boldsymbol{r},\omega)\rangle$ (and hence its influence on the field correlator $\langle\hat{\boldsymbol{E}}^{+}(\boldsymbol{r}',\omega)\hat{\boldsymbol{E}}^{-}(\boldsymbol{r},\omega)\rangle$), therefore, becomes direction-dependent. This imbalance produces a net momentum transfer parallel to the direction of motion, leading to a frictional force known as quantum friction [41–43].

Since the system exhibit optical gain, balanced with other dissipation channels, the system can be regarded as a kind of non-Hermitian systems, which can be characterised by PT symmetry [44,45]. Indeed, the relatively moving bodies may spontaneously breaks the PT symmetry [16,48,49]: If the velocity approaches a threshold (e.g., the typical critical velocity is $v_{\mathrm{c}} \approx 10^5\ \mathrm{m/s}$ for plasmonic slabs with a surface plasma frequency $\omega_{\mathrm{s}} = 2\pi \times 1.5\ \mathrm{THz}$ and the distance between the two slabs $d = 10\ \mathrm{nm}$ [49]), above which motion-induced gain dominates dissipation in the system, the force diverges, signalling the laser-type instability [44–49]. This validates the claim made in the previous section: the MQED framework based on the linear theory breaks down in the unstable regime.

(ii) *Hall-like (transverse) Forces*. —Not only the mechanical motion of a macroscopic body but also the motion of electric carriers inside the body, driven by a static electric bias, can modify fluctuations near the body. Several studies have shown that the longitudinal conductance of a conductor can give rise to frictional forces parallel to an applied bias [50–52], while the transverse conductance leads to unusual forces perpendicular to the bias [17].

Applying a bias to a medium with a Hall conductance, the fluctuation spectrum $\langle \hat{\boldsymbol{j}}_{\mathrm{N}}^{+}(\boldsymbol{r}',\omega)\hat{\boldsymbol{j}}_{\mathrm{N}}^{-}(\boldsymbol{r},\omega)\rangle$ acquires an asymmetry in the direction perpendicular to the bias. Thus, the field fluctuations are transversally "dragged", generating an imbalance transverse to the bias and hence a transverse force. This Hall-like fluctuation-induced force has no analogue in equilibrium passive systems and originates purely from the bias-induced modification of the electromagnetic correlations with a Hall conductance in low-symmetry materials [17].

## Conclusion

We have presented the MQED framework and its extension to active systems. We showed how the noise-current and field operators are phenomenologically constructed from a collection of harmonic oscillators representing material degrees of freedom. The central role of the field correlator was emphasised as the quantity underlying both radiative observables and mechanical effects. When optical gain is introduced, the electromagnetic fluctuations are modified, leading to unusual phenomena such as frictional and Hall-like forces.

## Acknowledgements

D.O. was supported by the JSPS Overseas Research Fellowship and is currently supported by the RIKEN special postdoctoral researcher program.